\documentclass[fleqn,twoside]{article} 
\usepackage{epsf,multicol,ifthen}
\usepackage{ujp}
\usepackage[cp1251]{inputenc}
\usepackage[english,russian]{babel}
\usepackage{latexsym, amssymb,amsmath}

\nazva{GENERALIZED DEFORMED OSCILLATORS IN FRAMEWORK OF UNIFIED
$(q;\alpha,\beta,\gamma;\nu)$-DEFORMATION AND THEIR OSCILLATOR
ALGEBRAS}

\udk{538.9; 538.915; 517.957}

\nazvacol{GENERALIZED DEFORMED OSCILLATORS IN FRAMEWORK OF UNIFIED
$(q;\alpha,\beta,\gamma;\nu)$- DEFORMATION}

\avtor{I.M. BURBAN}%
\avtorcol{I.M. BURBAN}%
\inst{Bogolyubov Institute for Theoretical Physics, Nat. Acad. of
Sci. of Ukraine}%
\adr{(14b, Metrolohichna Str., 03143 Kyiv, Ukraine)}%

\begin{document}           
\setcounter{page}{1}%

\newcommand {\be}{\begin{equation}}
\newcommand {\ee}{\end{equation}}
\newcommand{\bea}{\begin{eqnarray}}
\newcommand{\eea}{\end{eqnarray}}
\newcommand{\ba}{\begin{array}}
\newcommand{\ea}{\end{array}}

\newcommand{\Uqso}{{U'_q({\rm so}_n)}}


\maketitl                 

\begin{multicols}{2}
\anot{%
The aim of this paper is to review of our results on description
of the multi-parameter deformed oscillators and their oscillator
algebras. We define generalized $(q;\alpha,\beta,\gamma;\nu)$-
deformed oscillator algebra and study its irreducible
representations. The Arik-Coon oscillator with the main relation
$aa^+ - q a^+a = 1,$ where $q > 1$ is embedded in this framework.
We find connection of this  oscillator with the Askey
$q^{-1}-$Hermite polynomials. We construct family of the
generalized coherent states associated with these polynomials and
give their explicit expression in terms of standard special
functions.  By means of the solution of appropriate classical
Stielties moment problem we prove the (over)completeness relation
of these states.
}%

\section{Introduction}

\noindent The oscillator algebra plays a central role in the
investigation of many physical systems. It is also useful in the
theory of Lie algebra representations. The physical motivation of
the study of deformed boson and fermion quanta is connected with
the hope that the deformed oscillators in nonlinear systems will
play the same role as usual oscillator in the standard quantum
mechanics.

The investigation of the one-parameter deformed oscillator
algebras in theoretical physics originated from the study of the
dual resonance models of strong interactions \cite {AC}. The
$q$-deformed analog of the harmonic oscillator was introduced in
the well-known papers \cite{B,M}.

In parallel with the one-parameter deformed commutation relations
the two-parameter $(p,q)$-deformation of this relations has been
introduced \cite {CJ, J}. The connection $(p,q)$-deformed
oscillator algebra with $(p,q)$-hypergeometric functions has been
established in \cite{BK0}.The two-parameter deformed boson algebra
invariant under the quantum group $SU_{q_1/q_2}$ ( 'Fibonacci'
oscillator) was studied in \cite{AD}.

A wide class of the generalized deformed oscillator algebras
studied in literature is connected with the generalized deformed
oscillators. The attractive description of the systems of
particles with continuous interpolating (Bose and Einstein)
statistics, the theory of fractional quantum Hall effect, high
$T_c$ superconductivity require of an deformation of the canonical
commutation relations. The $q$-deformed oscillators are used
widely in the molecular and the nuclear spectroscopy. Nonlinear
vector coherent states (NVCSs) of $f$-deformed spin-orbit
Hamiltonians became the focus of attention the research of \cite
{GH}. This class includes multi-parameter generalization of the
one- and two- parameter deformed oscillator algebras \cite{ GH,
C-C-N-U, BDY, Bur, MLD, Bur1, Bur2}. Some of them have found
application in investigation of various physical systems.

The multi-parameter deformed quantum algebras was used in work
\cite {R} to construct integrable multi-parameter deformed quantum
spin chains. It is naturally, the magnification of the number of
deformation parameters makes the method of the deformations more
flexible. Although multi-parameter deformed quantum algebra in
some cases can be mapped onto standard one-parameter deformed
algebra \cite {CZ, P} the physical results in both cases are not
the same. The Hamiltonian of the electromagnetic monochromatic
field in the Kerr medium \cite {WM} is embedded in the framework
of four-parameter deformed oscillator algebra \cite {MLD}. This
gives the complete description of the energy spectrum of this
system. The most general famous examples of the multi-parameter
deformed oscillator algebras is the $(q;\alpha,\beta,\gamma)$- and
$(q,p;\alpha,\beta,l)$- deformations of the one- and two-
parameter deformed oscillator algebras \cite {C-C-N-U, BDY, Bur}.

The modified oscillator algebra \cite {V} has found applications
in the study of the integrability of the two-particle Calogero
model \cite {BHV}. This algebra has been generalized to
$C_{\lambda}$-extended oscillator algebra \cite{Q} with the hope
to exploit theirs for construction of new integrable models.
 The generalized $C_{\lambda}$-extended oscillator algebra - the
$S_N$-extended oscillator algebra, supplemented with a certain
projector- underlying an operator solution N-particle Calogero
model \cite{E}. For the same purpose a "hybrid" model of the
$q$-deformed and the modified oscillator algebras has been
proposed \cite{BEM}.

To complete the cycle of these ideas we have proposed the
generalized $(q;\alpha,\beta,\gamma;\nu)$-deformed oscillator
algebra as "the synthesis" of the
$(q;\alpha,\beta,\gamma)$-deformed \cite{B, C-C-N-U} and
$\nu$-modified oscillator algebras \cite {V}. The unified form of
the $(q;\alpha,\beta,\gamma;\nu)$-deformed oscillator algebra is
useful not only because its gives unified approach to the
well-known examples of the deformed oscillator algebras, but also
because it gives new partial examples of the deformed oscillators
with useful properties. By means of selection of special values of
deformation parameters we have separated a generalized deformed
oscillator connected with generalized discrete Hermite II
polynomials \cite {Bur1}. By theirs means we have constructed the
Barut-Girardello type coherent states of this oscillator. We have
found the conditions on the
$(q;\alpha,\beta,\gamma;\nu)$-deformation parameters at which the
$(q;\alpha,\beta,\gamma;\nu)$-deformed oscillator approximate the
usual anharmonic oscillator in the homogeneous Kerr medium. The
Arik-Coon oscillator with the main relation $aa^+ - q a^+a = 1,$
where $q > 1,$ is embedded in these framework. We find connection
of this  oscillator with the Askey $q^{-1}-$Hermite polynomials.
We construct family of the generalized coherent states, associated
with these polynomials, and give their explicit expression in
terms of standard special functions.  By means of the solution of
appropriate classical Stielties moment problem we prove the
(over)completeness relation of these states.

\section {Oscillator algebra and its generalized
deformations}

The oscillator algebra of the quantum harmonic oscillator is
defined by canonical commutation
relations\begin{equation}\label{burban: algeb}[a, a^+] = 1, [N, a]
= - a, [N, a^+]= a^+.
\end{equation}
 It allows for the different types of deformations. Some of them
have been called {\it generalized deformed oscillator algebras}
\cite{C-C-N-U, D1, D2,MMP}. Each of them defines an algebra
generated by the elements (generators) $\{{\bf 1},a, a^+, N \}$
and the relations \[\, a^+a = f(N),\, aa^+ = f(N+1),\]
\begin{equation}\label{burban: funct} [N, a]= - a,\,[N, a^+] = a^+,
\end{equation} where $f$ is called  {\it the structure
function of the deformation}. Among them - the multiparameter
generalization of one-parameter deformations
\cite{{C-C-N-U},{BDY},{Bur},{MLD},{Bur1},{Q},{BEM},{MMP}}.

Let us recount some of them.

 1. The Arik-Coon $q$-deformed oscillator algebra \cite{AC}
 \[ a a^+ - q a^+ a = 1,\,[N,a]= - a,\,[N, a^+] = a^+,\, q\in{\mathbb R}_+,
 \]\begin{equation}\label {burban: ar-coon} f(n )= \frac{1-q^n}{1-q}.\end{equation}
2. The Biedengarn--Macfarlane $q$-deformed oscillator algebra
\cite{B,M}\[ aa^+ - q a^+a = q^{- N},\, aa^+ - q^{-1} a^+a =
q^{N}\]\[
 [N, a]= - a,\, [N, a^+] = a^+,\]
 \begin{equation} f(n)=\frac{q^n - q^{-n}}{q - q^{-1}},\,q\in{\mathbb
 R}_+.\end{equation}
3. The Chung-- Chung--Nam--Um generalized
$(q;\alpha,\beta)$-deformed oscillator algebra \cite{C-C-N-U}
\[ aa^+ - q a^+a = q^{\alpha N + \beta},\,
[N, a]= - a,\, [N, a^+] = a^+,\, q\in{\mathbb
R}_+,\]\begin{equation} f(n)=
\begin{cases}
q^{\beta}\frac{q^{\alpha n} - q}{q^{\alpha} - q}, &\text {if
$\alpha \ne 1;$}\\ nq^{n-1 + \beta}, &\text {if $\alpha = 1,$}\\
\end{cases}
\end{equation}
where $\alpha, \beta \in{\mathbb R}.$

4. The generalized $(q;\alpha,\beta,\gamma)$-deformed oscillator
algebra \cite{BDY}
\[ aa^+ - q^{\gamma} a^+a = q^{\alpha N + \beta},\, [N, a]= - a,\,
[N, a^+] = a^+,\,\]\begin{equation}\label{burban: bor} f(n)=
\begin{cases}
q^{\beta}\frac{q^{\alpha n}- q^{\gamma n}}{q^{\alpha}- q^{\gamma
n}}, &\text {if $\alpha \ne \gamma ;$}\\ nq^{n-1 + \beta}, &\text
{if $\alpha = \gamma, $}\\
\end{cases}
\end{equation}
where $q\in{\mathbb R}_+,\alpha, \beta,\gamma \in{\mathbb R}.$

 5. The
$\nu$-modified oscillator algebra \cite {{V},{BHV}}
\[ [a, a^+]= 1 + 2\nu K,\,[N, a]= - a,\, [N, a^+] =
a^+,\]\[\, a K = - K a,\, a^+ K = - K a^+,\, K^2 = 1,\, \nu \in
\mathbb R,\]
\begin{equation}\label{burban: vas}
f(n)=
\begin{cases}
2k + 1 + 2\nu, &\text {if $ n = 2k ;$}\\ 2k+2, &\text {if $n =
2k+1$}.\\
\end{cases}
\end{equation}
This oscillator, as it has been shown in \cite{BHV}, is linked to
two-particle Calogero model \cite {Cal}.

6. The deformed $C_{\lambda}$-extended oscillator algebra \cite
{Q} is defined by the relations \[ [a, a^+]_q \equiv aa^+ - q a^+a
=\]\[ H(N) + K(N)\sum _{k = 0}^{\lambda -1}\nu_k P_k ,\quad [N,
a]= - a,\,[N, a^+] = a^+,\]\begin{equation}\label{burban: Ques} \,
a K = - K a,\, a^+ K = - K a^+,\, K^2 = 1,\, \nu_k \in \mathbb
R,\end{equation} where  $\nu_k\in \mathbb R$ and $H(K),\, K(N)$
are real analytic functions. This algebra permits the two Casimir
operators $C_1= e^{2\pi N} and \quad C_2 = \sum_{k = 0}^{\lambda
-1}e^{-2\pi i(N-k)/\lambda}P_k.$

7. The new $(q;\nu)$-deformed oscillator \cite {BEM} \[ aa^+ - q
a^+ a = (1 + 2\nu K) q^{-N},\, [N, a]= - a,\]\[ [N, a^+] = a^+,\,
K a = - a K,\,K a^+ = - a^+ K,\, K^2 = 1,\]\begin{equation} f(n)=
\Bigl(\frac{q^n - q^{-n}}{q - q^{-1}}+ 2\nu \frac {q^n - (-1)^n
q^{-n}}{q + q^{-1}}\Bigr)\end{equation} has been defined by the
combination of the idea of Biedenharn-- Macfarlane \cite {{B},{M}}
$q$-deformation with the Brink, Hanson and Vasiliev idea \cite
{BHV} of the $\nu$-modification of the oscillator algebra.

8. In order to complete this cycle of ideas we consider a
$(q;\alpha,\beta,\gamma;\nu)$-deformed oscillator algebra --
"hybrid" of the $(q;\alpha,\beta,\gamma)$-deformed (\ref {burban:
bor}) and the $\nu$-modified (\ref{burban: vas}) oscillator
algebras -- or, more exactly, an oscillator defined by the
generators $\{I, a, a^+, N, K\}$ and relations
\[ aa^+ - q^{\gamma} a^+ a = (1 + 2\nu K) q^{\alpha N +
\beta},\]\[[N, a]= - a,\, [N, a^+] = a^+ ,K a = - a ,
\]\begin{equation}\label{burban: first}   K a^+ = - a^+ K,\,
[N, K] = 0,\, N^+ = N,\, K^+ = K,
\end{equation}
where $q \in {\mathbb R}_+,\alpha,\beta \in {\mathbb R}, \nu \in
{\mathbb R}-\{0\}.$ This model unifies all deformations 1. - 7. of
the oscillator algebra (\ref{burban: algeb}).

\section{Generalized $(q;\alpha,\beta,\gamma;\nu)$-deformed
oscillator algebra and its simplest properties}

{\it (a) $(q;\alpha,\beta,\gamma;\nu)$-deformed structure
function.} Description of an deformed oscillator algebra requires
the determination of the deformation structure function $f(n).$

Equations (\ref{burban: funct}) and (\ref{burban: first}) imply
the recurrence relation
\begin{equation}\label{burban: struct0} f(n+1) - q^{\gamma} f(n) =
\Bigl(1 + 2 \nu(-1)^n\Bigr)q^{\alpha n + \beta}. \end{equation}
Its solution is obtained by the mathematical induction method
\cite {Bur0}. The solution of the equation (\ref{burban: struct0})
with the initial value $f(0)= 0$ is given by the following formula
\[ f(n) =\]
\begin{equation} \label{burban: struct}
\begin{cases}
q^{\beta}\Bigl(\frac {q^{\gamma n} - q^{\alpha n}} {q^{\gamma} -
q^{\alpha}} + 2\nu \frac {q^{\gamma n} - (-1)^n q^{\alpha n}}
{q^{\gamma} + q^{\alpha}}\Bigr),&\text{if $\alpha \ne \gamma;$}\\
n q^{\gamma(n-1) + \beta} + 2 \nu q^{\gamma(n-1) + \beta}
\Bigl(\frac{1 - (-1)^n}{2}\Bigr),&\text{ if $\alpha =
\gamma.$}\end{cases}
\end{equation}

{\it (b) Useful formulas.} Following formulas will be useful for
the study of this algebra. One of them is
\begin{equation}\label{burban: useful} a(a^+)^n - q^{\gamma n }\, (a^+)^n a =
[n;\alpha,\gamma;\nu K](a^+)^{n-1}q^{\alpha
N+\beta},\end{equation} where $n\ge 1,$ and the other one
\[[n;\alpha,\gamma;\nu K] = \]
\begin{equation}
\begin{cases}
\Bigl(\frac {q^{\gamma n} - q^{\alpha n}} {q^{\gamma} -
q^{\alpha}}+ 2\nu K\frac {q^{\gamma n} - (-1)^n q^{\alpha n}}
{q^{\gamma} + q^{\alpha}}\Bigr),&\text{if $\alpha \ne \gamma$;}\\
n q^{\alpha(n-1)} + 2\nu K q^{\alpha(n-1)} \Bigl(\frac{1 -
(-1)^n}{2}\Bigr),&\text{ if $\alpha = \gamma$}\end{cases}
\end{equation}
is deduced by the method of mathematical induction.
 The direct
calculations leads to (\ref{burban: useful}). For
$[n;\alpha,\gamma;\nu K]$ the second formula gives the generating
function $\sum_{n = 0}^{\infty}[n;\alpha,\gamma;\nu K]z^n =$
\begin{equation}
\begin{cases}
\frac{z}{1-q^{\gamma}z}\Bigl(\frac{1}{1-q^{\alpha}z}+2\nu K
\frac{1}{1+q^{\alpha}z}\Bigr) , &\text {if $\alpha \ne \gamma;$}\\
\frac{z}{(1-q^{\gamma}z)^2}+ 2\nu K\frac{z}{1-q^{2\gamma}z^2} ,
&\text {if $\alpha = \gamma.$}\\
\end{cases}
\end{equation}
{\it (d) Deformed $C_2$-extended and
$(q;\alpha,\beta,\gamma;\nu)$-deformed oscillator algebras.}

The defining relations of the deformed $C_2$-extended oscillator
are given by
\[aa^+ - q^{\gamma} a^+ a = H(N)+ \nu\Bigl( E(N+1) + q^{\gamma}
E(N)\Bigr)(P_0 - P_1),\]
\[ [N, a^+]= a^+,[N,P_{k}] = 0, a^+ P_{k} = P_{k
+1}a^+,\]
\begin{equation}\label{burban: deform}P_1 + P_2 = I,P_{k} P_{l} = \delta_{k,l}P_{l},
\end{equation}
where $q, \nu \in {\mathbb R},k,l =1, 2,$ and $E(N), H(N)$ are
real analytic functions. As we saw above the deformed extended
oscillator algebra $C_{\lambda}$ permits the two Casimir operators
$C_1, C_2.$ In case of the $C_2$-extended oscillator algebra they
have the form
\begin{equation}\label{burban: casimir} C_1 = e^{2\pi i N},\quad C_2 =
e^{i\pi N}K.
\end{equation}
Let us define the operator \begin{equation}\label{burban: kasimir}
{\tilde C}_3 = q^{-\gamma N}\Bigl(D(N) + \nu E(N) K -
a^+a\Bigr),\end{equation} where $D(N),\, E(N)$ are some analytic
functions of $N$. The operator ${\tilde C}_3$ will be the Casimir
operator of the oscillator algebra (\ref{burban: deform}) if the
only one condition $[{\tilde C}_3, a] = 0$ holds.  It amounts to
determination of the solution of the equations
\[ K(N)\nu_{k}=\quad E(N+1)\beta_{k+1}-
q^{\gamma}E(N)\beta_{k},\]
\begin{equation}\label{burban: addi}
H(N) = D(N+1)- q^{\gamma}D(N),
\end{equation}
where $\nu_0 = - \nu_1 =\nu, \beta_0 = 0,\beta_2 = 0, \beta_1 =
\nu,\quad k = 0, 1.$ Substituting the solution $ E(N) = 2
q^{\alpha N + \beta}/({q^{\gamma} + q^{\alpha}})$ of the equation
of (\ref{burban: addi}) and $ H(N) = q^{\alpha N + \beta}$ in
(\ref{burban: deform}), we obtain the commutation relations of the
$ (q;\alpha,\beta,\gamma;\nu)$-deformed oscillator algebra
(\ref{burban: first}). Moreover, the solution $$ D(N) =
\begin{cases} q^{\beta}\Bigl(\frac{q^{\gamma N } - q^{\alpha N
}}{q^{\gamma} - q^{\alpha}} + 2\nu\frac{q^{\gamma N}}{q^{\gamma} +
q^{\alpha}}\Bigr),&\text{if $\gamma \ne \alpha
$;}\\q^{\beta}(q^{\gamma(N-1)}N + \nu q^{-\gamma}),&\text{if
$\gamma =\alpha $}\end{cases} $$ of the first equation
(\ref{burban: addi}) gives the explicit form of the Casimir
operator
\[{\tilde C}_3 =\]
\[
\begin {cases} q^{-\gamma N }\Bigl((\frac{q^{\gamma
N} - q^{\alpha N }}{q^{\gamma} - q^{\alpha}} + 2 \nu
\frac{q^{\gamma N} - (-1)^N q^{\alpha N}}{q^{\gamma} +
q^{\alpha}})q^{\beta} - a^+a \Bigr),&\\\text {if
$\alpha\ne\gamma;$}\\ q^{-\gamma N}\Bigl(N + \nu(1 +
(-1)^N)q^{\gamma N + \beta}- a^+a\Bigr),&\\\text{if $\alpha =
\gamma $}.\end{cases}\]

\section {Classification of representations of unified
$(q;\alpha,\beta,\gamma;\nu)$-deformed oscillator algebra}

As has been shown in the previous Section the
$(q;\alpha,\beta,\gamma;\nu)$-deformed oscillator algebra allows
for a nontrivial center what means that it has irreducible
non-equivalent representations \cite {Rid, Ch}. We give a
classification of these representations by a method similar to the
one in the articles \cite{Q1, KMM}.

Due to the relations (\ref{burban: first}) and (\ref {burban:
casimir}) there exists a vector $|0\rangle$ such
that\[a^+a|0\rangle = \lambda_0 |0\rangle, aa^+ |0\rangle =
\mu_0|0\rangle,\]\begin{equation} N |0\rangle = \varkappa_0
|0\rangle, K|0\rangle = \omega e^{-i\pi
\varkappa_0}|0\rangle,\end{equation} where $\langle 0|0 \rangle =
1$ and $\omega$ is the value of the Casimir operator $C_2$ in the
given irreducible representation. By means of (\ref{burban:
useful}) we find that vectors
\begin{equation}
|n\rangle' = \begin{cases} (a^+)^n|0\rangle, &\text{if $n\ge 0
;$}\\ (a)^{-n}|0\rangle, &\text{if $n < 0$}\\
\end{cases}
\end{equation}
are eigenvectors of the operators $a^+a$ and $aa^+:$
\begin{equation}\label{burban: comun} a^+ a|n\rangle' = \lambda_n |n\rangle',
\quad aa^+ |n\rangle' = \mu_n |n\rangle'.\end{equation}
 Let us define new
system of the orthonormal vectors $\{|n\rangle\}_{n = - \infty}^{n
= \infty},$ by
\begin{equation}
|n\rangle = \begin{cases}\Bigl(\prod_{k = 1}^{n} \lambda_k
\Bigr)^{-1/2} (a^+)^n|0\rangle, &\text{if $n\ge 0 ;$}\\
\Bigl(\prod_{k = 1}^{-n}\lambda_{n+k}\Bigr)^{-1/2}
(a)^{-n}|0\rangle, &\text{if $n < 0$}.\\
\end{cases}
\end{equation}
Then the relations (\ref{burban: first}) are represented by the
operators\[a^+|n\rangle = \sqrt{\lambda_{n+1}}|n+1\rangle, \quad a
|n\rangle = \sqrt{\lambda_n}|n-1\rangle, \]
\begin{equation}\label{burban: rep} N|n\rangle = (\varkappa_0 +
n)|n\rangle, \quad K |n\rangle =
\frac{(-1)^n}{2\nu}B|n\rangle,\end{equation} where $B = 2 \nu
\omega e^{-i\pi \varkappa_0}\in {\mathbb R}.$ Due to
non-negativity of the operators $a^+a,$ $aa^+$ we have
$\lambda_n\ge 0$ and $\mu_n\ge 0.$

From the identity $a(a^+a)|n\rangle = (aa^+)a|n\rangle$ we find
\begin{equation}\label{burban: relat}\lambda_n = \mu_{n-1}\end{equation}
and from (\ref{burban: rep}) the recurrence relation
\begin{equation}\label{burban: resur}
\lambda_{n+1} - q^{\gamma}\lambda_n = \Bigl(1 + (-1)^n B
\Bigr)q^{\alpha(n +\varkappa_0)+ \beta}.
\end{equation} Take into account the relation (\ref{burban: first}) the
solution of equation (\ref{burban: resur}) can be represented by
\[\lambda_n = \]\begin{equation} \label{burban: solu}
\begin{cases} \lambda_0 q^{\gamma n} +
q^{\alpha\varkappa_0 + \beta}\Bigl(\frac {q^{\gamma n} - q^{\alpha
n}} {q^{\gamma} - q^{\alpha}} + B \frac {q^{\gamma n} - (-1)^n
q^{\alpha n}} {q^{\gamma} + q^{\alpha}}\Bigr),&\text{if $\alpha
\ne \gamma$;}\\\lambda_0 q^{\gamma n} + q^{\gamma \varkappa_0 +
\beta }\Bigl(nq^{\gamma(n-1)} + B \frac{1 - (-1)^n}{2}
\Bigr),&\text{ if $\alpha = \gamma.$}
\end{cases}\end{equation}
The nonnegativity of $\lambda_n$\, ($\gamma - \alpha \ne 0$)
implies for $n = 2k$ and for $n = 2k+1$ respectively
\[\Bigl(\lambda_0 q^{-(\alpha \varkappa_0 + \beta)}+
\frac{1}{q^{\gamma}- q^{\alpha}} +\]
\begin{equation}\label{burban: noneq0} \frac{B}{q^{\gamma} +
q^{\alpha}}\Bigr) \ge q^{-2(\gamma
-\alpha)k}\Bigl(\frac{1}{q^{\gamma} - q^{\alpha}} + \frac{
B}{q^{\gamma} + q^{\alpha}} \Bigr), \end {equation}
\[\Bigl(\lambda_0 q^{-(\alpha \varkappa_0 + \beta)}+
\frac{1}{q^{\gamma}- q^{\alpha}} +\]
\begin{equation}\label{burban: noneq}
\frac{B}{q^{\gamma} + q^{\alpha}}\Bigr) \ge q^{-(\gamma
-\alpha)(2k+1)}\Bigl(\frac{1}{q^{\gamma} - q^{\alpha}} - \frac{
B}{q^{\gamma} + q^{\alpha}} \Bigr).
\end{equation}
The representations of the generalized oscillator algebra are
reduced to the four classes of unireps:

(i) Assume $q <1, \alpha =\gamma > 0,$ or $q >1, \alpha =\gamma >
0, B < 0.$ The nonnegativity $\lambda_n$ implies
\[\lambda_0 q^{-\gamma(\kappa_0 +1)-\beta}(n +
q^{-\gamma(n-1))}B\frac{1-(-1)^n}{2})\ge 0.\] Therefore there
exists $n_0$ such that $\lambda_n < 0 $for all $n < n_0.$ After
possible re-numbering we may assume \[a|0\rangle = 0, \,\lambda_0
= 0.\] Therefore the representation of the relations (\ref{burban:
first}) is given by formula (\ref{burban: rep}) with \[ \lambda_n
= q^{\gamma \varkappa_0 + \beta }\Bigl(nq^{\gamma(n-1)} + B
\frac{1 - (-1)^n}{2} \Bigr), \,\forall n\ge 0.\]

(ii) Assume $\gamma - \alpha > 0,$ $q > 1$ $(\gamma-\alpha < 0, 0
< q < 1).$  From this it follows that at least one of the numbers
$\frac{1}{q^{\gamma} - q^{\alpha}} \pm \frac{ B}{q^{\gamma} +
q^{\alpha}}$ is positive. Due to (\ref {burban: noneq0}),(\ref
{burban: noneq}) there exists $n_0$ such that for all even or odd
$n < n_0, \lambda_n < 0$  and after possible renumbering we may
assume
\[ a| 0 \rangle = 0, \lambda_0 = 0.\]
The nonnegativity condition for $\lambda_n$ implies $B \ge -1$
gives:
\begin{itemize}
\item If $ B > -1.$ The representations relations (\ref{burban: first})
are given by formulae
(\ref{burban: rep}) with
\begin{equation}\label{burban: second} \lambda_n =
q^{-\alpha \varkappa_0 + \beta }\Bigl(\frac{q^{\gamma n
}-q^{\alpha n}}{q^{\gamma}-q^{\alpha}} - \frac{q^{\gamma n} -
(-1)^n q^{\alpha n}}{q^{\gamma} + q^{\alpha}}\Bigr).
\end{equation}
The arbitrary values of the parameter $\varkappa_0,$ and $B > -1$
define nonequivalent infinite-dimensional
representations(\ref{burban: rep}) of of the relation
(\ref{burban: first}).
\item If $B = -1.$ In this case due to (\ref{burban: relat})$\lambda_1 = \mu_0$
and the representations (\ref{burban: rep}) are defined by
\begin{equation}\label{burban: dim1}
a = a^+ = 0, \quad N = \varkappa_0,\quad K = - \frac{1}{2\nu}.
\end{equation}
\end{itemize}
(iii) Assume $q < 1,$ $\gamma - \alpha > 0,$  $(q
> 1,\gamma-\alpha < 0 ).$
From this it follows that at least one and only one of the numbers
$\frac{1}{q^{\gamma} - q^{\alpha}} \pm \frac{ B}{q^{\gamma} +
q^{\alpha}}$ is positive. Due to (\ref{burban: noneq0}),
(\ref{burban: noneq}) there exists $n_0$ such that for $n
> n_0$ the $\lambda_n$ is negative for even and odd values $n.$
This implies $a^+|n\rangle = 0$ for some $n\ge n_0.$ After
possible renumering we have
\begin{equation}
a^+|0\rangle = 0
\end{equation}
This condition implies $\lambda_1 = 0,$ or $\lambda_0 = -
q^{\alpha\varkappa_0 + \beta - \gamma}(1+B).$  The condition
$\lambda_0\ge 0$ is equivalent to $B \le -1.$

\noindent If $B = -1,$ one obtains the representation
(\ref{burban: dim1}).

\noindent If $B < -1,$ it leads to
\[\lambda_n = q^{\alpha \varkappa_0 + \beta + \gamma
n} \Bigl(\frac{1- q^{(\alpha - \gamma)n}}{q^{\gamma} - q^{\alpha}}
- q^{-\gamma}\Bigr) +\]
\begin{equation}\label{burban: solu2}
B \Bigl(\frac{1- q^{(\alpha -\gamma)n}(-1)^n}{q^{\gamma} +
q^{\alpha}} - q^{-\gamma} \Bigr)\ge 0.
\end{equation}
The nonnegativity condition for $\lambda_n$ gives a restriction
for possible values of $B:$
\begin{itemize}
\item For values $B < \frac{q^{\gamma}+ q^{\alpha}}{q^{\gamma}+
q^{\alpha}}$ we have $\lambda_n > 0.$ Therefore the representation
of (\ref{burban: first}) with $\lambda_n $ (\ref {burban: solu2})
gives representations (\ref{burban first}). The arbitrary values
of parameter $\varkappa_0$ and $ B < \frac{q^{\gamma}+
q^{\alpha}}{q^{\gamma}+ q^{\alpha}}$ distinguish irreducible
representation.

\item For values $B = \frac{q^{\gamma}+ q^{\alpha}}{q^{\gamma}+
q^{\alpha}}$ we have $\lambda_n > 0.$ The vector space of this
representation is a span of the two-dimensional vectors $$ \left(
\begin{array}{c}
\psi_{-1} \\ \psi_0
\end{array}\right)
$$ and due to (\ref{burban: rep}) the representation are defined
by $$ a = \left(
\begin{array}{cc}
0&\sqrt{\frac{2 q^{\alpha\varkappa_0 + \beta}}{q^{\alpha} -
q^{\gamma}}}\\ 0& 0
\end{array}\right),
a^+ = \left(
\begin{array}{cc}
0& 0\\ \sqrt{\frac{2 q^{\alpha\varkappa_0 + \beta}}{q^{\alpha} -
q^{\gamma}}}& 0
\end{array}\right), $$
\begin{equation}
N = \left(
\begin{array}{cc}\chi_0 - 1
& 0\\0& \chi_0
\end{array}\right), K = \frac{1}{2\nu}\frac{q^{\gamma}+
q^{\alpha}}{q^{\alpha} - q^{\gamma}}\left(
\begin{array}{cc}
1&0\\0& 1
\end{array}\right).
\end{equation}
These representations are distinguished by the arbitrary values
$\varkappa_0,$ and $B = \pm\frac{q^{\gamma} +
q^{\alpha}}{q^{\gamma}- q^{\alpha}},$ $\lambda_0 =
\frac{2q^{\alpha(\varkappa_0 ) + \beta}}{q^{\alpha} -
q^{\gamma}}.$
\end{itemize}
(iv) Let us assume $q < 1,$ $\gamma - \alpha > 0,$ $(q >
1,\gamma-\alpha < 0)$ and $\lambda_n$ be defined by (\ref{burban:
solu}). This and the conditions that both values
$\frac{1}{q^{\gamma} - q^{\alpha}} \pm \frac{ B}{q^{\gamma} +
q^{\alpha}}$ are nonpositive (then at last one of them must be
strictly negative) lead to cases (see (\ref {burban:
noneq0}),\,(\ref {burban: noneq})). There are following
possibility:
\begin{equation} a)\qquad\Bigl(\lambda_0 q^{-(\alpha \varkappa_0 +
\beta)} + \frac{1}{q^{\gamma}- q^{\alpha}} + \frac{B}{q^{\gamma} +
q^{\alpha}}\Bigr) < 0.
\end{equation}
Due to (\ref{burban: noneq0}),(\ref{burban: noneq}) there exists
$n_0$ such that for $n
> n_0$ the $\lambda_n$ is negative for even and odd values $n.$
This implies as in (ii):
\begin{itemize}
\item $B < -\frac {q^{\gamma} + q^{\alpha}}{q^{\gamma}-
q^{\alpha}}.$ These representations of the relations (\ref{burban:
first}) are given by formulae (\ref{burban: rep}) with  $\lambda_n
$ (\ref {burban: second}).
\item $-1 < B.$ These representations of the relations(\ref{burban:
first}) are given by formulae (\ref{burban: dim1}).
\end{itemize}
\begin{equation} b)\qquad \Bigl(\lambda_0 q^{-(\alpha
\varkappa_0 + \beta)}+ \frac{1}{q^{\gamma}- q^{\alpha}} +
\frac{B}{q^{\gamma} + q^{\alpha}}\Bigr) > 0\end{equation} This
condition implies $\lambda_n > 0, \forall n\in \mathbb{Z}.$ This
representation is given by formulae (\ref {burban: rep}) with
$\lambda_n$ as (\ref{burban: solu}) for $\alpha\ne\beta$ and $n\in
\mathbb Z.$
\begin{equation} c)\qquad \Bigl(\lambda_0 q^{-(\alpha \varkappa_0 +
\beta)}+ \frac{1}{q^{\gamma}- q^{\alpha}} + \frac{B}{q^{\gamma} +
q^{\alpha}}\Bigr) = 0. \end{equation}
\begin{itemize}
\item $|B| < -\frac {q^{\gamma} + q^{\alpha}}{q^{\gamma}-
q^{\alpha}}$ implies $\lambda_n > 0 \forall n\in {\mathbb Z}.$ The
representations the same as in b).
\item
$|B| = -\frac {q^{\gamma} + q^{\alpha}}{q^{\gamma}- q^{\alpha}}$
implies $\lambda_n > 0 \forall n\in\mathbb{Z}.$ It follows
$\lambda_n = 0,\forall n =2k .$ The vector space of this
representation spanned by the two-dimensional vectors \[ \left(
\begin{array}{c}
\psi_0 \\ \psi_{1}
\end{array}\right).
\]
Therefore the representation is two-dimensional and given by the
formula \[ a = \left(
\begin{array}{cc} 0&0\\\sqrt{\frac{q^{\alpha(\varkappa_0+1) + \beta}}{q^ {\gamma}
- q^{\alpha}}}& 0
\end{array}\right),
a^+ = \left(
\begin{array}{cc}
 0& \sqrt{\frac{q^{\alpha(\varkappa_0+1) + \beta}}{q^
{\alpha} - q^{\gamma}}}\\0& 0
\end{array}\right),
\]
\begin{equation}
N = \left(
\begin{array}{cc}
\chi_0& 0\\0& \chi_0 + 1
\end{array}\right), K = \frac{1}{2\nu}\frac{q^{\gamma}+
q^{\alpha}}{q^{\gamma} - q^{\alpha}}\left(
\begin{array}{cc}
-1&0\\0& 1
\end{array}\right).\qquad
\end{equation}
These representations are defined by arbitrary values of
$\varkappa_0$, and  $\lambda_0 = 0,$ $B = - \frac {q^{\gamma} +
q^{\alpha}}{q^{\gamma}- q^{\alpha}}$.
\end{itemize}

\section{Generalized $(q;\alpha,\beta,\gamma;\nu)$-deformed oscillators
and nonlinear quantum optical model} \label{sec:six}

In this Section we study some aspects concerning the possible
interpretation of $(q;\alpha,\beta,\gamma;\nu)$-deformed
non-interacting systems describing non-deformed interacting
systems.
 We consider an anharmonic oscillator in quantum optics defined
by the Hamiltonian $H$ to describe laser light in a nonlinear Kerr
medium. In lower order it is of the form \cite{WM}
\begin{equation}\label{burban: kerrmed}
H_{Kerr} = \frac{\hbar\omega_0}{2} (2N+1) + \frac{\kappa}{2}
N(N-1),
\end{equation}
where $\kappa$ is the real constant related to nonlinear
susceptibility $\chi^{3}$ of the Kerr medium. In the framework of
the $(q;\alpha,\beta, \gamma;\nu)$-deformed oscillator algebra
with the help of a corresponding choice of the deformation
parameters we shall construct operators approximating this
Hamiltonian.

If $\alpha = \gamma$ we consider the $(q;\alpha,\beta,
\gamma;\nu)$-deformed oscillator algebra (\ref {burban: first})
and define the corresponding Hamiltonian by
\begin{equation} H =
\frac{\hbar\omega_0}{2}(a^+a + aa^+),\end{equation} or
\[H_N = \frac{\hbar\omega_0}{2} [q^{\gamma (N-1) + \beta} \Bigl( N +  2 \nu
(\frac{1 - (-1)^N}{2}\Bigr) +
\]\begin{equation}\label{burban: tamm}   q^{\gamma N +
\beta}\Bigl( N+1 + 2 \nu \frac{1 - (-1)^(N+1)}{2}\Bigr)].
\end{equation}
Assuming small values of $\gamma$ and $\beta$ in this operator we
obtain an approximation of this Hamiltonian \[H_N
=\frac{\hbar\omega_0}{2} [2N+I+ 2\gamma(N-I)N +(2\gamma + 2 \beta
+ 2 \nu)N +\]\begin{equation}\label{burban: approx}
 O(\gamma^2,\beta^2,\beta\nu,\gamma \nu)].
\end{equation}
Comparing (\ref{burban: kerrmed}) and (\ref {burban: approx}) we
obtain their equivalence if
\[\gamma = \frac{\kappa}{2\hbar\omega_0},\quad \beta + \nu =
 - \frac{\kappa}{2\hbar\omega_0}.\]If $\gamma\ne \alpha\,\quad \nu = 0$ we consider $(q;\alpha,\beta,
\gamma;\nu)$-deformed oscillator algebra (\ref {burban: first})
and the Hamiltonian
\begin{equation} H =
\frac{\hbar\omega_0}{2}aa^+,\end{equation} or
\begin{equation}\label{burban: ham}
H_N = \frac{\hbar \omega_0}{2} \frac{q^{\alpha(N+1)}-
q^{\gamma(N+1)}}{q^{\alpha}- q^{\gamma}}
\end{equation}
If we introduce the new deformation parameters $q = e, \alpha =
\rho + \mu, \gamma =\rho -\mu, \beta = 0,$ then Hamiltonian
(\ref{burban: ham}) takes the form
\begin{equation}
H_N = e^{\rho N}\frac{\hbar\omega_0}{2}\frac{e^{\mu(N+1)}-
e^{-\mu(N+1)}}{e^{\mu}-e^{-\mu}} =
\end{equation}
\begin{equation}
e^{\rho N}\frac{\hbar\omega_0}{2}(e^{\mu N}+ e^{\mu (N-1)}+\ldots
+ e^{-\mu( N-1)}+ e^{-\mu N}).\end{equation} Assuming small values
of $\mu$ and $\rho$ in this operator and using the expansion
\noindent \[ e^{\mu N} = I + \mu N + \frac{\mu^2}{2} N^2+ \ldots
;\]
\[\vdots\]
\noindent\[e^{-\mu N} = I - \mu N + \frac{\mu^2}{2}N^2-\ldots
\]
\[e^{\rho N}\simeq I+\rho N + \ldots \] we obtain
\[\frac{e^{\mu(N+I)}-e^{-\mu(N
+I)}}{e^{\mu}-e^{-\mu}}\simeq (2N + I) +
\frac{\mu^2}{2}\frac{N(N+1)(2N+1)}{6},\] and
\[ H_N = \frac{\hbar\omega_0}{2}[(2N+1)+ (\frac{\mu^2}{2} + 2
\rho)N(N-1)+(\frac{2}{3}\mu^2 + 3\rho)N
+\]\begin{equation}\label{burban: ost} O(
\rho^2,\rho\mu^2,\mu^4)].
\end{equation}
Comparing (\ref{burban: kerrmed}) and (\ref{burban: ost}) we
obtain their equivalence if
\begin{equation}\label {burban: ker}\mu^2 = -\frac{
9}{2}\rho, \quad \rho = -\frac{2\kappa}{\hbar\omega_0}.
\end{equation}

\section {Generalized $(q,p;\alpha,\beta,l)$-deformed oscillator}

We introduce the multi-parameter generalization of the two-
parameter deformed oscillator algebra \cite {Bur}
$(p,q;\alpha,\beta,l)$-deformed canonical commutation relations by
the formulas \[ aa^+ - q^{l} a^+ a = p^{-\alpha N-\beta}, aa^+ -
p^{-l}a^+ a = q ^{\alpha N + \beta}
\]
\begin{equation}\label{burban: rel3a}[N, a] = - \frac{l}{\alpha} \,a, \quad[N, a^+] =
\frac{l}{\alpha }a^+.\end{equation} It is easy to see that
function $f(n)$ for this case has the form
\begin{equation}
f(n)= \frac{p^{-\alpha n - \beta }- q^{\alpha n + \beta}}{p^{-l }-
q^{l}}
\end{equation}
with $\alpha,\,\beta,\in R, l\in Z .$ The creation and
annihilation operators $a,\,a^+$ and the operator $N$ of the
relations (\ref{burban: rel3a}) act on the Hilbert space ${\cal
H}$ with the basis $\{|n\rangle\}, n = 0,1,2 \ldots$ as follows
\[a^+\,|n\rangle =\left(\frac{p^{-\alpha -\beta -l}-q^{\alpha +
\beta + l }}{p^{-l}-q^{l}}\right)^{1/2}|n+l/\alpha\rangle\]
\begin{equation}a|n\rangle = \left(\frac{p^{-\alpha -\beta} -
q^{\alpha + \beta}}{p^{- l }-q^{l}}\right)^{1/2}|n-l/\alpha\rangle
.\end{equation} We define the difference operator (the Jackson
derivative)
\begin{equation}
D f(z)= \frac{f(p^{-\alpha}z)p^{- \beta} -
f(q^{\alpha}z)q^{\beta}}{(p^{-l}-
q^{l})z^{{l}/\alpha}},\end{equation} where $f(z)$ belong to a
space of functions (analytic if $l/\alpha$ is an integer). It
follows $$ Dz^n = \frac{z^n}{z^{{l}/\alpha}}\, \frac{p^{- \alpha n
- \beta}- q^{\alpha n + \beta}}{p^{-l} -
q^{l}}\,\frac{1}{(n)!}\frac{d^{n}z^{n}}{d z^{n}}.$$ If $l/\alpha $
is an integer, then for an analytic function  $ f(z)=
\sum_{n=0}^{\infty}a_n z^n $ we have
\begin{equation} D f(z) = \sum_{n=1}^{\infty}\frac{z^n}
{z^{l/\alpha}}\frac{p^{-\alpha n - \beta} - q^{\alpha n +
\beta}}{p^{-l}- q^{l}}\frac{1}{n!}\frac{d^n}{dz^n}f(z).
\end{equation}
Then in this space we can give the "coordinate" realization of the
relations (\ref{burban: rel3a}):
\[q^N: f\to q^{z\frac{d}{d z}}f = f(q z),
p^{-N}: f\to p^{-z\frac{d}{d z}}f = f(p^{-1}z),
\]
\begin{equation}\label{burban: an4}a: f\to D f,  a^+ :f\to z^{{l}/\alpha}f, N: f\to
z\frac{d}{d z}.
\end{equation}
Indeed, from  (\ref{burban: an4}) we obtain
\[ N a^+ f(z) = z \frac{d}{d z}(z^{l/\alpha} f(z)) =
l/\alpha z^{{l}/\alpha}f + z^{1 + l/\alpha}\frac{d}{d z} f(z),\]\[
a^+ N f(z) = l/\alpha z^{1+l/\alpha}\frac{d}{d z}f(z).\] It
follows
\begin{equation} [N, a^+]f = l/\alpha\,a^+ f.\end{equation} and analogously,
\begin{equation}[N, a]= -{l/\alpha}\,a. \end{equation}

In a similar way, from (\ref{burban: an4})we obtain
\[a^+ \,a f(z) =
\frac{f(p^{-{\alpha}}z)p^{-\beta}- f(q^{\alpha}
z)q^{\beta}}{p^{-l} - q^{l}},\]
\begin{equation}    a\,a^+ f(z) = \frac{f(p^{-\alpha}z)p^{-l-\beta}-
f(q^{\alpha} z)q^{{l} + \beta}}{p^{-{l}}-q^{l}}.\end{equation}
and, therefore, representation of the relations (\ref {burban:
rel3a}).

\section {Generalized $(q;a,b,c;0)$-deformed oscillator}

The unified form of the $(q;\alpha,\beta,\gamma;\nu)$-deformed
oscillator algebra is useful not only because its gives unified
approach to well-know deformed oscillator algebras 1.-5. of Sec.
2., but also because it gives new partial examples of the deformed
oscillators with useful properties.

Let us consider the example of such oscillator algebra. It is
convenient to introduce in (\ref {burban: struct}) the new
deformation parameters
\begin{equation}\label {burban: exch}\alpha = 2 a + c - 1,\quad \beta = 2 a + b,
\quad \gamma = 2 a,\end{equation} and assume $\nu = 0,\alpha \ne
\gamma.$ Then  we obtain generalized deformed oscillator
\[ [N, a]= -a,\quad [N, a^+] = a^+,\]
\begin{equation}
\label{burban: grelation2} a a^+ - q^{2a} a^+ a = q^{2 a (N + 1) +
b}q'^N,
\end{equation}
with the structure function of deformation \[f(n)= [n;q;a,b,c;0] =
q^{2a n + b}\Bigl( \frac{1-q^{(c-1)n}}{1-q^{(c-1)}}\Bigr)\]
\begin{equation}\label{burban:struc}  = q^{2a n + b}\Bigl(
\frac{1-q'^{ n}}{1-q'}\Bigr),\quad q'= q^{c-1},\end{equation} and
whose properties we shall study below.

\section { Arik-Coon oscillator with $q > 1$ and $(q; a, b, c ;
0)$-deformation }

Fixing the values of the parameters in (\ref{burban: grelation2})
we arrive to the oscillators well studied in literature: $a = 1/2,
b = -1, c = 0$ (the Arik-Coon oscillator with  $q < 1$) connected
with the Rogers $q-$ Hermite polynomials, $a = - 1, b = 2, c = 2$
(the oscillator, connected with the discrete $q$-Hermite II
polynomials \cite{BD}). The replacement $q \to 1/q$ in
(\ref{burban: ar-coon}) leads to the oscillator \begin{equation}
[N, a]= - a,\quad [N, a^+] = a^+, a a^+ - q^{-1} a^+ a =
1,\end{equation} where $q < 1$ which is equivalent to the
oscillator (\ref {burban: grelation2}), where $a = - 1/2, b = 1, c
= 2,$ with the structure function of the deformation
\begin{equation}\label{burban: struct}  f(n)= [n; q; -1/2, 1, 2 ;
0] = q^{- n + 1}\Bigl( \frac{1-q^n}{1-q} \Bigr), q < 1
\end{equation} connected \cite {BK} with $q^{-1}$-Hermite
polynomials Askey \cite {A}.

As has shown \cite{BK} the operator $Q = a^+ + a,$ or
\begin{equation}\label{burban: coo} Q|n\rangle = r_n|n+1\rangle +
r_{n-1}|n-1\rangle,\end{equation} where
\[ r_n =
[n+1; q; a, b, c; 0]^{1/2} =
q^{-n/2}\Bigl(\frac{1-q^{n+1}}{1-q}\Bigr)^{1/2}\] is a unbounded
symmetric operator. Its closure ${\bar Q}$ is not self-adjoint
operator and has the deficiency indices (1,1) \cite {Ber}.
Defining the generalized eigenfunction $Q |x\rangle = x|x\rangle
$, where $|x\rangle = \sum_{n = 0}^{\infty} P_n(x)|n\rangle,$ we
obtain the recurrence relation
\begin{equation}\label{burban: jac}
r_{n-1} P_{n-1}(x) + r_n P_{n+1}(x) = x P_n(x).
\end{equation}
The coefficients ${P}_n(x;q)$ of this equation satisfy the
relation
\[x P_n(x;q) =\]
\[ q^{1/2}(1-q)^{-1/2} \Bigl(
{q^{-(n+1)}}(1-q^{n+1})\Bigr)^{1/2}{P}_{n+1}(x;q) +\]
\begin{equation}
\label{burban: re}q^{1/2}(1 -q)^{-1/2}\Bigl({q^{-n}} (1-
q^n)\Bigr)^{1/2}{ P}_{n-1}(x;q).
\end{equation}
Introducing the change of variables $2y = q^{-1/2}(1-q)^{1/2}x,$
$\psi(x;a)= P(2q^{1/2}(1-q)^{-1/2}x) $ and
\begin{equation}\label{burban: bas}{\psi}_n(x;q) =
\frac{h_n(x;q)}{q^{-n(n+1)/4}(q;q)_n^{1/2}}
 \end{equation} we obtain the recurrence relation
\begin{equation}\label{burban: dublrel} 2x
h_{n}(x;q) = h_{n+1}(x;q)+ q^{- n }(1-q^{n})h_{n-1}(x;q) .
\end{equation}
The solution of this equation with initial conditions $ h_{0}(x;q)
= 1 ,\quad h_1(x;q) = 2x$ is given  by the $q^{-1}-$ Hermite
polynomials \cite {IM}\[ h_n(x;q) = \sum_{k = 0}^n
\frac{(q;q)_n}{(q;q)_k(q;q)_{n-k}}\times
\]
\begin{equation}\label{burban: her} (-1)^k
q^{k(k-n)} (x + \sqrt{x^2 + 1})^{n-k}.
\end{equation}
The orthogonality relation for these polynomials is
\begin{equation}\label{burban: ort}
\int_{-\infty}^{\infty}h_{m}(x; q)h_{n}(x; q)d\nu(x) =
q^{-n(n+1)/2}(q; q)_n \delta_{m,n}.
\end{equation}
(a) {\it Generalized Barut-Girardello coherent states.} We denote
by ${\cal H}_F$ the Hilbert space spanned by the basis vectors
$|n> = \psi_n(x;q),\quad n = 1, 2 ,\ldots $ of the orthogonal
polynomials (\ref{burban: bas}). We consider ${\cal H }_F $ as the
Fock space for the operators $a^+, a.$ These operators
(\ref{burban: grelation2}) in the space ${\cal H}_F $ are
represented as
\begin{equation} a|n\rangle = r_{n-1}|n-1\rangle,a^+|n\rangle = r_n |n+1\rangle
\end{equation}
The coherent states of the Barut-Girardello type for this
oscillator in the Fock space ${\cal H}_F $ are defined as
eigenvectors of the annihilation operator $a$ \begin{equation}
a\,|z\rangle = z\,|z\rangle,\quad z\in \mathbb{C}. \end{equation}
Thy are given by the formula
\begin{equation}\label{burban: def} |z\rangle = {\cal
N}^{-1}(|z|^2)\sum_{n=1}^{\infty}\frac{z^n}{r_{n-1}!}|n\rangle,
\end{equation} where ${\cal N}$ is normalized factor and
\[ r_{n}! =  \begin {cases} 1,&\text {if}\quad n = 0;
\\ r_{n}r_{n-1}\ldots r_{1},&\text{if}\quad n = 1, 2, \ldots .
\end{cases}\] We consider the coherent
states of this oscillator, connected with $q^{-1}$-Hermite
polynomials (\ref{burban: her}). They are given by the expression
({\ref{burban: def}}), where
\[ r_{n-1}! = \Bigl(\frac{q}{1-q} \Bigr)^{n/2}
q^{-n(n + 1)/4}(q;q)^{1/2}_n \] and
\[ |n\rangle = \psi_n(x;q)
= \frac{h_n(x;q)}{q^{-n(n+1)/4}(q;q)_n^{1/2}}.\] It follows
\[|z\rangle = {\cal N}^{-1}(|z|^2)|z\rangle \times\]
\[\sum_{n = 0}^{\infty}\frac{z^n \Bigl(
\frac{1-q}{q}\Bigr)^{n/2}}{q^{- n(n +
1)/4}(q;q)^{1/2}_n}{h}_n(x;q)
\frac{1}{q^{-n(n+1)/4}(q;q)_n^{1/2}},\] or
\begin{equation}\label{burban: prel} |z\rangle = {\cal N}^{-1}(|z|^2)\sum_{n =
0}^{\infty}(\sqrt{1-q})^n q^{n^2/2}\frac{h_n (x;q)
}{(q;q)_n}z^n.\end{equation} Take into account the relation
(\ref{burban: ort}) the normalizing factor can be written\[{\cal
N}^{2}(|z|^2) = \]
\begin{equation}\sum_{n =
0}^{\infty} \frac{q^{n(n-1)/2}}{(q;q)_n}(1-q)^{n}|z|^{2n} =
(-(1-q)|z|^2; q)_{\infty},\end{equation} or \begin{equation}{\cal
N}^{2}(|z|^2) = {}_0\phi_0(-; -q; -(1-q)|z|^2).\end{equation}
Using the generating function \cite {Koe}
\[\sum_{n =
0}^{\infty} \frac{t^n q^{n(n-1)/2}}{(q;q)_n}h_n(x;q)=\]
\begin{equation}\Bigl(- t(x + \sqrt{x^2 + 1}); t(\sqrt{x^2 + 1} -
x)\Bigr)_{\infty}\end{equation} for polynomials $h_n(x;q)$ and
(\ref{burban: prel}) one obtains
\[|z\rangle =\]
\begin{equation}
\frac{\Bigl(-z\sqrt{q(1-q)}(x + \sqrt{x^2 + 1});
z\sqrt{q(1-q)}(\sqrt{x^2 + 1} -
x)\Bigr)_{\infty}}{\Bigl(-(1-q)|z|^2;q\Bigr)_{\infty}^{1/2}}.
\end{equation}
(b){\it Completeness of generalized coherent states.} It is
necessary to prove the decomposition of unity formula
\begin{equation}\label{burban: compl}
\int \int_{\mathbb {C}}{\hat W}(|z|^2)|z\rangle\langle z|d^2 z =
\sum_{n = 0}^{\infty}|n\rangle\langle n|= 1,
\end{equation}
i.e., to construct a measure
\begin{equation}\label{burban: mea}
d\mu(|z|^2)= {\hat W}(|z|^2)d^2z,\, d^2 z =({\rm R}e z)({\rm I}m
z).
\end{equation}
Using (\ref{burban: def}) the relation (\ref{burban: compl}) can
be represented as
\begin{equation}
\sum_{n=0}^{\infty} \frac{\pi}{r_{n-1}^2}\int_0^{\infty} d x x^n
\frac{\hat W(x)}{{\cal N}^2(x)}|n\rangle\langle n| = 1, \,(x =
|z|^2).
\end{equation}
Defining
\begin{equation}
{\tilde W}(x)= \pi \frac{{\hat W(x)}}{{\cal N}^2(x)}
\end{equation}
we arrive to the solving of the classical moment problem
\begin{equation}\label{burban: mo}
\int_0^{\infty}d x x^n{\tilde W}(x) = r_{n-1}^2!=
\Bigl(\frac{1}{1-q}\Bigr)^n q^{-n(n-1)/2}(q;q)_n.
\end{equation}
and replacement of the variables ${W}(y)= \frac{1}{1-q} {\tilde
W}(\frac{y}{1-q})$ leads (\ref{burban: mo}) to the form
\begin{equation}\label{burban: moment}\int_0^{\infty}d y y^n{W}(y) =
q^{-n(n-1)/2}(q;q)_n.
\end{equation}
In order to solve moment problem (\ref{burban: moment}) we define
a q-exponential function \[e_q(x)={}_0\phi_0(x;q) =\sum_{n =
0}^{\infty}\frac{x^n}{c_{n-1}^2!} = \]
\begin{equation} \sum_{n =
0}^{\infty}\frac{x^n}{q^{-n+1}!(1-q^n)!}=
\sum_{n=0}^{\infty}\frac{q^{n(n-1)/2}}{(q;q)_n}x^n,
\end{equation} where
\cite{GR}
\begin{equation} {}_0\phi_0(x;q)= \Bigl(\begin
{array}{c c|c} 0&0&\\ -&{-}&\end{array} q;-x\Bigr) =
(-x;q)_{\infty}.\end{equation} Define deformed derivative by
\begin{equation}\label{burban: der}
[\frac{d}{d x}]_q f(x)= \frac{f(q^{-1}x)- f(x)}{q^{-1}x}.
\end{equation}
one obtain
\begin{equation} [\frac{d}{d x}]_q x^n =
\begin {cases} c_{n-1}^2 x^{n-1}, &\text {if}\quad n \ge 0;
\\ 0, &\text {if} \quad n = 0
\end{cases}
\end{equation}
and therefore
\begin{equation}
[\frac{d}{d x}]_q e_q(x)= e_q(x).
\end{equation}
The following Leibniz rule holds for this deformed derivative
\begin{equation}
[\frac{d}{d x}]_q [u(x)\cdot v(x)] =  \begin {cases} [\frac{d}{d
x}]_q u(x)\cdot v(q^{-1} x)+ u(x)\cdot [\frac{d}{d x}]_q v(x),
\\v(x) \cdot [\frac{d}{d x}]_q u(x)+ u(q^{-1}x)\cdot\frac{d}{d x}]_q
v(x).
\end{cases}
\end{equation}
From this rules and relation $e_q^{-1}(x)e_q(x) = 1$ we
obtain\[\frac{d}{d x}]_q \Bigl(e_q (x)e_q^{-1}(x)\Bigr) =\]
\begin{equation}
 \frac{d}{d
x}]_q e_q(x)\cdot e_q^{-1}(q^{-1}x)+ e_q(x)\cdot[\frac{d}{d x}]_q
e_q^{-1}(x) = 0,
\end{equation}
i.e.,
\begin{equation}
[\frac{d}{d x}]_q e_q^{-1}(x)= - e_q^{-1}(q^{-1}x).
\end{equation}
We now introduce the Jackson integral corresponding to the
derivative (\ref{burban: der})
\begin{equation}\label{burban: in}
\int_0^{\infty} f(t)dt_q = q^{-1}\sum_{l = 0}^{\infty} q^{-l+1}
f(q^{-l+1})+ q^{l+2} f(q^{l+2}).
\end{equation}
The formula integration by parts has the form
\[\int_0^{\infty}u(x)\cdot [\frac{d}{d x}]_q v(x) =\]
\begin{equation}\label{burban: bypar}
\int_0^{\infty}[\frac{d}{d x}]_q [u(x)\cdot v(x)]-\int_0^{\infty}
[\frac{d}{d x}]_q u(x)\cdot v(q^{-1} x).
\end{equation}
Let us consider the integral
\begin{equation}
I_n = \int_0^{\infty} x^n[\frac{d}{d x}]_q e_q^{-1}(x) = -
\int_0^{\infty} x^n e_q^{-1}(q^{-1}x).
\end{equation}
Using the formula (\ref{burban: bypar}) we obtain
\begin{equation}
I_n = \int_0^{\infty} x^n e_q^{-1}(q^{-1}x) = \int_0^{\infty}
x^{n-1} e_q^{-1}(q^{-1}x)c_{n-1}^2
\end{equation}
i.e.,
\begin{equation}\label{burban: val}
I_n = I_{n-1}c_{n-1}^2 \qquad n\ge 0,
\end{equation} or
\begin{equation}
I_n = c_{n-1}^2! = q^{- n(n-1)/2}(q,q)_{n}.
\end{equation}
Take into account (\ref {burban: in}) and (\ref {burban: val}) we
obtain solution of the classical moment problem (\ref{burban:
moment})\[W(y)=\]
\begin{equation}
q^{-1}\sum_{l = 0}^{\infty} y\Bigl[ \delta (y - q^{-l+1}) +
\delta(y-q^{l+1})\Bigr]e_q(q^{-1}y).
\end{equation}
so that\[{\bar W}(x) = \frac{1-q}{q}e_q
\Bigl(q^{-1}(1-q)x\Bigr)\times\]
\begin{equation}
\sum_{l=0}^{\infty}x\Bigl[\delta \Bigl( (1-q)x - q^{-l+1}\Bigr) +
\delta\Bigl((1-q)x-q^{l+1}\Bigr)\Bigr].
\end{equation}
The measure in (\ref{burban: mea}) is given by
\[d \mu(|z|^2) = \frac{1-q}{\pi q}\Bigl(-(1-q)|z|^2; q)_{\infty}\Bigr)
e_q\Bigl(q^{-1}(1-q)|z|^2\times\]
\begin{equation}
\sum_{l = 0}^{\infty}|z|^2\Bigl[\delta \Bigl((1-q)|z|^2 -
q^{-l+1}\Bigr) + \delta\Bigl((1-q)|z|^2- q^{l+1}\Bigr)\Bigr].
\end{equation}

\section {Conclusions}
The aim of this article is observation of the results obtained in
\cite {Bur, Bur1, Bur2, Bur0} on the generalized
$(q;\alpha,\beta,\gamma;\nu)$-deformed oscillator algebra. We
study general properties of this algebra. The Arik-Coon oscillator
with the main relation $aa^+ - q a^+a = 1,$ where $q > 1,$ is
embedded in the framework of the unified
$(q;\alpha,\beta,\gamma;\nu)$-deformed oscillator algebra. In
addition we discuss for this last case uniqueness the solution of
the Stielties moment problem. The subsequent investigations of the
properties of the $(q;\alpha,\beta,\gamma;\nu)$-deformed
oscillator algebra and its applications can be found in the works
\cite{GH, GH1, KO, AB, HNB, GKR, GRe}.


The research was partially supported by the special Program of
Division of Physics and Astronomy of the National Academy of
Science of Ukraine.

{}

\rezume{%
╙╟└├└╦▄═┼═▓ ─┼╘╬╨╠╬┬└═▓ ╬╤╓╚─╦▀╥╬╨╚  ┬ ╨└╠╩└╒
╬┴"к─═└═╬п $(q;\alpha,\beta,\gamma;\nu)$- ─┼╘╬╨╠└╓│┐ ▓ п╒
╬╤╓╚╦╦▀╥╬╨═▓ └╦├┼┴╨╚} {╚.╠. ┴єЁсрэ} {

╠хЄю■ Ў│║┐ ёЄрЄЄ│ ║ юуы ф эр°шї Ёхчєы№ЄрЄ│т ч яюсєфютш
єчруры№эхэшї $(q;\alpha,\beta,\gamma;\nu)$-фхЇюЁьютрэшї
юёЎшы ЄюЁiт │ ┐ї юёЎшы ЄюЁэшї рыухсЁ. ╠ш тштўр║ью ┐┐ эхчт│фэ│
яЁхфёЄртыхээ . ╟юъЁхьр, юёЎшы ЄюЁ └Ё│ър-╩єэр │ч уюыютэшь
ёя│тт│фэю°хээ ь $aa^+ - q a^+a = 1,$ where $q > 1,$ тъырфр║Є№ё  т
Ў│ Ёрьъш. ╠ш чэрїюфшью чт" чюъ Ў№юую юёЎшы ЄюЁр ч хЁь│Єютшьш
$q^{-1}$-фхЇюЁьютрэшьш яюы│эюьрьш └ёъ│. ╠ш сєфє║ью ё│ь"■
ъюухЁхэЄэшї ёЄрэ│т Єшяє ┴рЁєЄр-─ц│ЁрЁфхыыю фы  Ў№юую юёЎшы ЄюЁр.
╟р фюяюьюую■ Ёючт" чъє т│фяют│фэю┐ ъырёшўээю┐ яЁюсыхьш ьюьхэЄ│р
╤Є│ы№Є║ёр ьш фютюфшью тырёЄшт│ёЄ№ (яхЁхяютэхэюёЄ│)яютэюЄш Ўшї
ёЄрэ│т.
 }

\rezume{%
╬┴╬┴┘┼══█┼ ─┼╘╬╨╠╚╨╬┬└══█┼ ╬╤╓╚╦╦▀╥╬╨█ ┬ ╨└╠╩└╒ ╬┴┌┼─╚═┼══╬╔
$(q;\alpha,\beta,\gamma;\nu)$-─┼╘╬╨╠└╓╚╚ ╚ ╚╒ ╬╤╓╚╦╦▀╥╬╨═█┼
└╦├┼┴╨█ } {╚.╠. ┴єЁсрэ} {

╓хы№■ ¤Єющ ёЄрЄ№ш  ты хЄё  юсючЁхэшх эр°шї Ёхчєы№ЄрЄют яюёЄЁюхэш 
юсюс∙хэ√ї $(q;\alpha,\beta,\gamma;\nu)$-фхЇюЁьшЁютрээ√ї
юёЎшыы ЄюЁют ш шї юёЎшыы ЄюЁэшї рыухсЁ. ╠√ шчєўрхь хх эхяЁштюфшь√х
яЁхфёЄртыхэш . ┬ўрёЄэюёЄш , юёЎшыы ЄюЁ └Ёшър-╩єэр шч уыртэшь
ёююЄэю°хэшхь $aa^+ - q a^+a = 1,$ уфх $q
> 1,$ тъырф√трхЄё  т ¤Єш Ёрьъш. ╠√ эрїюфшь ёт ч№ ¤Єюую
юёЎшыы ЄюЁр ё  $q^{-1}$-¤ЁьшЄютшьш яюышэюьрьш └ёъш. ╠ш эрїюфшь
ёхьхщёЄтю ъюухЁхэЄэ√ї ёюёЄю эшщ Єшяр ┴рЁєЄр-─цшЁрЁфхыыю фы  ¤Єюую
юёЎшы ЄюЁр. ╧Ёш  яюью∙ш Ёх°хэш  ёююЄтхёЄтє■∙хщ ъырёёшўхёъющ
яЁюсыхьь√ ьюьхэЄют ╤ЄшыЄ№хёр ь√ фюърч√трхь ётющёЄтю
(яхЁхяюыэхээюёЄш)яюыэюЄ√ ¤Єшї ёюёЄю эшщ. }

\end{multicols}

\end{document}